\documentclass[twocolumn,pra,showpacs,superscriptaddress]{revtex4}
\usepackage{amssymb}
\usepackage{amsmath}
\usepackage{graphicx}
\usepackage{subfigure}
\usepackage{natbib}
\usepackage{epsfig}
\usepackage{amsfonts}
\usepackage{mathrsfs}
\usepackage{CJK}
\usepackage[toc,page,title,titletoc,header]{appendix}

\begin{document}
\title{Role of initial system-bath correlation on coherence trapping}

\author{Ying-Jie Zhang}
 \affiliation{Shandong Provincial Key
Laboratory of Laser Polarization and Information Technology,
Department of Physics, Qufu Normal University, Qufu 273165, China}
 \affiliation{Beijing National Laboratory of Condensed Matter Physics, Institute of
Physics, Chinese Academy of Sciences, Beijing 100190, China}

\author{Wei Han}
 \affiliation{Shandong Provincial Key
Laboratory of Laser Polarization and Information Technology,
Department of Physics, Qufu Normal University, Qufu 273165, China}

\author{Yun-Jie Xia}
\email{yjxia@mail.qfnu.edu.cn}
 \affiliation{Shandong Provincial Key
Laboratory of Laser Polarization and Information Technology,
Department of Physics, Qufu Normal University, Qufu 273165, China}

\author{Yan-Mei Yu}
 \affiliation{Beijing National Laboratory of Condensed Matter Physics, Institute of Physics,
Chinese Academy of Sciences, Beijing 100190, China}

\author{Heng Fan}
\email{hfan@iphy.ac.cn}
 \affiliation{Beijing National Laboratory of Condensed Matter Physics, Institute of Physics,
Chinese Academy of Sciences, Beijing 100190, China}
\affiliation{Collaborative Innovation Center of Quantum Matter,
Beijing 100190, China}

\date{\today}
\begin{abstract}
We study the coherence trapping of
a qubit correlated initially with a non-Markovian bath in a pure dephasing
channel. By considering the initial qubit-bath
correlation and the bath spectral density, we find that the initial qubit-bath correlation can lead to a more
efficient coherence trapping than that of the initially
separable qubit-bath state. The stationary coherence in the long
time limit can be maximized by optimizing the parameters of the
initially correlated qubit-bath state and the bath spectral density.
In addition, the effects of this initial correlation on the maximal
evolution speed for the qubit trapped to its stationary coherence
state are also explored.

\end{abstract}
\pacs {03.65.Ta, 03.65.Yz, 03.67.-a, 42.50.-p}

\maketitle

{\it{Introduction.}}
Quantum state takes the form of superposition which leads to quantum
coherence. Quantum coherence plays a
central role in the applications of quantum physics and quantum
information science \cite{1,2}.
However, it is fragile due to interactions of the
environment. Understanding of
quantum coherence dynamics of an open system is a very important
task in many areas of physics ranging from quantum optics to quantum
information processing. It is known that many quantum open systems exhibit
non-Markovian behavior with a flow of information from the
environment back to the system \cite{3,4,5,6,7}. This presence of
non-Markovian effects can induce the long-lasting coherence in
biological surroundings and the steady state entanglement in
coherently coupled dimer systems \cite{8,9}. By considering the pure
dephasing non-Markovian bath, decay of quantum coherence of the
system would be terminated in a finite time, such that the system can
partly retain coherence in the long time limit. This new phenomenon,
known as $coherence$ $trapping$ \cite{10}, is important for quantum
information processing since the effective long-time quantum
coherence of the system is preserved. Coherence trapping of a
quantum system is mainly related to the open dynamics, and is
generally analyzed in the fact that the system and bath are
initially separable.
As is well known, however, the
initial system-bath correlations are important for the
the dynamics of the open systems.
The distinguishability of quantum states would increase
in the presence of initial system-bath correlations
\cite{16,17}. The information flow between the system and
its bath and the corresponding degree of non-Markovianity can also
be influenced by these initial correlations \cite{18,19,20,21}.
On the other hand, the standard master equation
approach to open systems may not be appropriate unless a product
state is explicitly prepared \cite{11,12,13,14,15,015}.
So the coherence trapping of an open system due to the initial
system-bath correlations should be studied both
physically and methodologically.

In this paper, we will concentrate on the following questions:
 how do the initial system-bath correlations affect
coherence trapping of the system? which form of the initially
correlated system-bath state can maximize the stationary coherence
of the system? We consider the pure dephasing model of a qubit
initially correlated with a zero-temperature Ohmic-Like bath.
We will show that the initial qubit-bath correlation can lead
to the more efficient coherence trapping, while the lower initial
coherence of the qubit is induced by this initial correlation.
In the long time limit, the stationary coherence of the qubit can be
maximized by choosing the optimal parameters of the initially
correlated qubit-bath state and the optimal Ohmicity parameter of
the bath.

Furthermore, the task to drive an initial state to a prescribed
 target state in the shortest possible time is significant
for quantum control  in many areas of physics, such as quantum
 computation \cite{021}, fast population transfer in quantum optics \cite{0021}, and
 quantum optimal control protocols \cite{022,0022}.
 This minimum evolution time, which is defined as quantum
speed limit (QSL) time
\cite{22,23,24,25,26,27,28,29,30,31,32,33,34,35}, is a key method in
characterizing the maximal speed of evolution of quantum systems.
Here in order to speed up the evolution from an initial coherence
state to its stationary coherence state, we further focus on the
interactions of the initial qubit-bath correlated state, the
spectral density function of the bath and the QSL time. Remarkably,
we find that the initial qubit-bath correlation can reduce the
QSL time for the occurrence of coherence trapping. The maximal
evolution speed for the qubit trapped to its stationary coherence
state can also be controlled by optimizing the parameters of the
initial qubit-bath correlated state and the bath spectral density
function.

{\it{Model.}}
 Let us consider an exactly solvable model,
 in which the process of energy dissipation is negligible
and only pure depahsing is a mechanism for decoherence of the qubit.
The associated Hamiltonian reads (setting $\hbar=1$),
\begin{eqnarray}
H=\omega_{0}\sigma_{z}+\int^{\infty}_{0}{\omega}a^{\dag}_{\omega}a_{\omega}d\omega+\int^{\infty}_{0}\sigma_{z}[g_{\omega}a^{\dag}_{\omega}+g^{*}_{\omega}a_{\omega}]d\omega,\label{1}
\end{eqnarray}
where the operator $\sigma_{z}$ is defined by
$\sigma_{z}=|e\rangle{\langle}e|-|g\rangle{\langle}g|$, associated
with the upper level $|e\rangle$ and the lower level $|g\rangle$ of
the qubit; $a_{\omega}$ and $a^{\dag}_{\omega}$ are the bosonic
annihilation and creation operators for the bath, which is
characterized by the frequency $\omega$; $g_{\omega}$ is the
coupling constant of the interactions of the qubit with the bath,
and $g^{*}_{\omega}$ is the complex conjugate to $g_{\omega}$. The
Hamiltonian in Eq. (\ref{1}) can be rewritten in the block-diagonal
structure \cite{38,39} $H=diag[H_{e},H_{g}]$, where
$H_{e/g}=\pm\omega_{0}+\int^{\infty}_{0}{\omega}a^{\dag}_{\omega}a_{\omega}d\omega\pm\int^{\infty}_{0}[g_{\omega}a^{\dag}_{\omega}+g^{*}_{\omega}a_{\omega}]d\omega$.

Here, we consider the situation where a correlated initial state of
the qubit-bath system in the form \cite{17},
\begin{eqnarray}
|\Psi(0)\rangle=c_{e}|e\rangle\otimes|\xi_{0}\rangle+c_{g}|g\rangle\otimes|\xi_{\lambda}\rangle,\label{2}
\end{eqnarray}
with the non-zero complex numbers $c_{g/e}$ are satisfied
$|c_{e}|^{2}+|c_{g}|^{2}=1$. And we assume that $|\xi_{0}\rangle$ is
a bath ground state and
$|\xi_{\lambda}\rangle=C^{-1}_{\lambda}[(1-\lambda)|\xi_{0}\rangle+\lambda|\xi_{f}\rangle]$
is a bath superposition state of the ground state $|\xi_{0}\rangle$
and a coherent state $|\xi_{f}\rangle=D(f)|\xi_{0}\rangle$. The
displacement operator $D(f)$ reads
$D(f)=\exp\{\int^{\infty}_{0}[f_{\omega}a^{\dag}_{\omega}-f^{*}_{\omega}a_{\omega}]d\omega\}$
for an arbitrary square-integrable function $f$. The constant
$C_{\lambda}=\sqrt{(1-\lambda)^{2}+\lambda^{2}+2\lambda(1-\lambda)Re\langle\xi_{0}|\xi_{f}\rangle}$
normalizes the state $|\xi_{\lambda}\rangle$, where $Re$ is a real
part of $\langle\xi_{0}|\xi_{f}\rangle$ in the bath Hilbert space.
The correlation parameter $\lambda\in[0,1]$ determines the initial
correlation of the qubit and bath. Through performing the
Hamiltonian described in Eq. (\ref{1}), the state of the total
system at any time $t$ is given by $
|\Psi(t)\rangle=c_{e}|e\rangle\otimes|\psi_{e}(t)\rangle+c_{g}|g\rangle\otimes|\psi_{g}(t)\rangle,$
where $|\psi_{e}(t)\rangle=exp(-iH_{e}t)|\xi_{0}\rangle$ and
$|\psi_{g}(t)\rangle=exp(-iH_{g}t)|\xi_{\lambda}\rangle$. Then the
reduced density matrix $\rho_{\lambda}(t)$ of the qubit at time $t$
reads, $\rho_{ee}(t)=|c_{e}|^{2}$, $\rho_{gg}(t)=|c_{g}|^{2}$ and
$\rho_{eg}(t)=
\rho^{*}_{ge}(t)=c_{e}c^{*}_{g}\Upsilon_{\lambda}(t)$, with the
dephasing rate $\Upsilon_{\lambda}(t)$.

The qubit dynamics is closely dictated by the spectral density
function characterising the qubit-bath interaction. In the following
the bath can be described by the family of Ohmic-Like spectra
$|g_{\omega}|^{2}=\alpha\omega^{\mu+1}\exp(-\omega/\omega_{c})$,
with $\omega_{c}$ being the cutoff frequency and $\alpha>0$ a
dimensionless coupling constant. By changing the $\mu$-parameter,
one goes from sub-Ohmic baths ($-1<\mu<0$) to Ohmic ($\mu=0$) and
super-Ohmic ($\mu>0$) baths, respectively. Furthermore, the coherent
state $|\xi_{f}\rangle$ can be calculated by the spectral density
function
$|f_{\omega}|^{2}=\omega^{\upsilon+1}\exp(-\omega/\omega_{c})$, with
$\upsilon>0$. So the initial state of the qubit-bath system can be
controlled by the parameters $\lambda$ and $\upsilon$. For the case
$\lambda=0$ the qubit and the bath are initially uncorrelated, the
dephasing rate can be obtained,
$\Upsilon_{0}(t)=\exp[-2i\omega_{0}t-r(t)].$ While for
$0<\lambda\leq1$ the initial correlation exists in the qubit-bath
system, one also finds,
$\Upsilon_{\lambda}(t)=C^{-1}_{\lambda}\{1-\lambda
+\lambda\exp[-2i\phi(t)+k(t)]\}\exp[-2i\omega_{0}t-r(t)]$, where,
\begin{eqnarray}
r(t)&=&4\alpha\Gamma[\mu]\omega^{\mu}_{c}\{1-\frac{\cos[\mu\arctan(\omega_{c}t)]}{(1+\omega^{2}_{c}t^{2})^{\mu/2}}\},\nonumber\\
k(t)&=&2\sqrt{\alpha}\Gamma[\vartheta]\omega^{\vartheta}_{c}\{1-\frac{\cos[\vartheta\arctan(\omega_{c}t)]}{(1+\omega^{2}_{c}t^{2})^{\vartheta/2}}\}-\frac{1}{2}\Gamma[\upsilon]\omega^{\upsilon}_{c},\nonumber\\
\phi(t)&=&\sqrt{\alpha}\Gamma[\vartheta]\omega^{\vartheta}_{c}\frac{\sin[\vartheta\arctan(\omega_{c}t)]}{(1+\omega^{2}_{c}t^{2})^{\vartheta/2}},
\label{4}
\end{eqnarray}
where $\Gamma[\cdot]$ is the Euler gamma function and the parameter
$\vartheta=(\mu+\upsilon)/2$.

{\it{Coherence trapping for the qubit.}}
 How to quantify quantum coherence of a quantum system now
becomes paramountly important. In recent years, a wide variety of
measures of coherence have been proposed \cite{40,41,42}. Currently,
Baumgratz, Cramer and Plenio find that the relative entropy of
coherence \cite{40},
\begin{eqnarray}
C(\rho)=S(\rho_{diag})-S(\rho),\label{5}
\end{eqnarray}
where $S(\rho)$ is the von Neumann entropy and $\rho_{diag}$ denotes
the state obtained from $\rho$ by deleting all off-diagonal
elements, and the intuitive $l_{1}$ norm of coherence,
\begin{eqnarray}
C_{l_{1}}(\rho)=\sum_{i,j,i{\neq}j}|\rho_{ij}|,\label{6}
\end{eqnarray}
are both general and proper measures of coherence. So in the
following we will choose the relative entropy of coherence $C(\rho)$
to measure the quantum coherence of the reduced density matrix
$\rho_{\lambda}(t)$ of the qubit in the presence of qubit-bath
initial correlation.

 If there is no correlations in the initial qubit-bath state, the qubit dephasing
$\Upsilon_{0}(t)$ is characterized by exponential decay of the qubit
coherence, hence will predict vanishing coherence in the long time
limit in the Ohmic and sub-Ohmic dephasing baths \cite{10}. While
for the super-Ohmic baths, the qubit dephasing will stop after a
finite time, therefore leading to coherence trapping. This behavior
can realize the effective long-time coherence protection. In the
following, we would mainly see the effect of the initially
correlated qubit-bath state on coherence trapping of a qubit in the
super-Ohmic bath model. The preparation of this initially correlated
qubit-bath state can be obtained by non-local operations with two
steps \cite{17}. Firstly, one prepares the bath state
$|\xi_{\lambda}\rangle$, and then products it with the ground state
of the qubit $|g\rangle$. Then, one would superpose
$|g\rangle\otimes|\xi_{\lambda}\rangle$ with the product state
$|e\rangle\otimes|\xi_{0}\rangle$, with the weights $c_{g/e}$,
respectively. And the initial correlations of the qubit-bath system
can be controlled by the parameters $c_{g/e}$, $\lambda$ and the
function $f_{\omega}$.
\begin{figure}[tbh]
\includegraphics*[bb=0 0 298 275,width=4cm, clip]{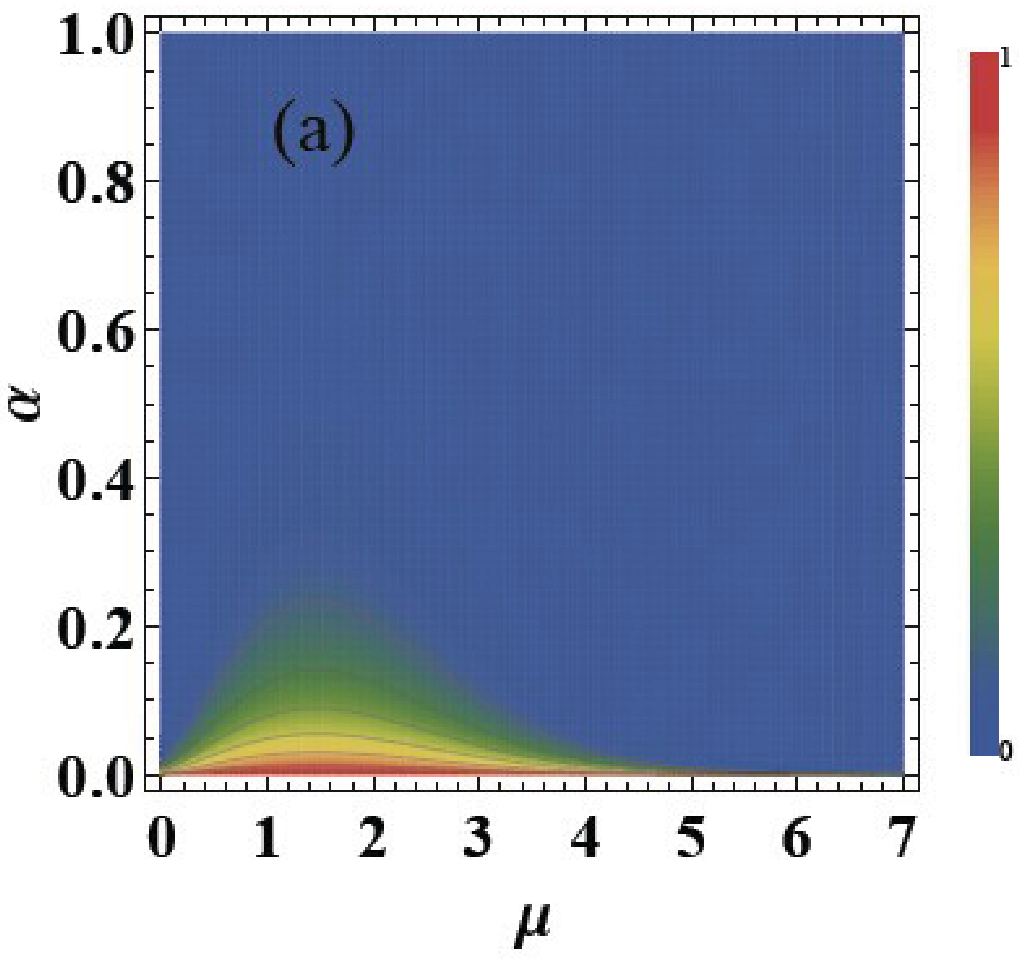}
\includegraphics*[bb=0 0 306 275,width=4.1cm, clip]{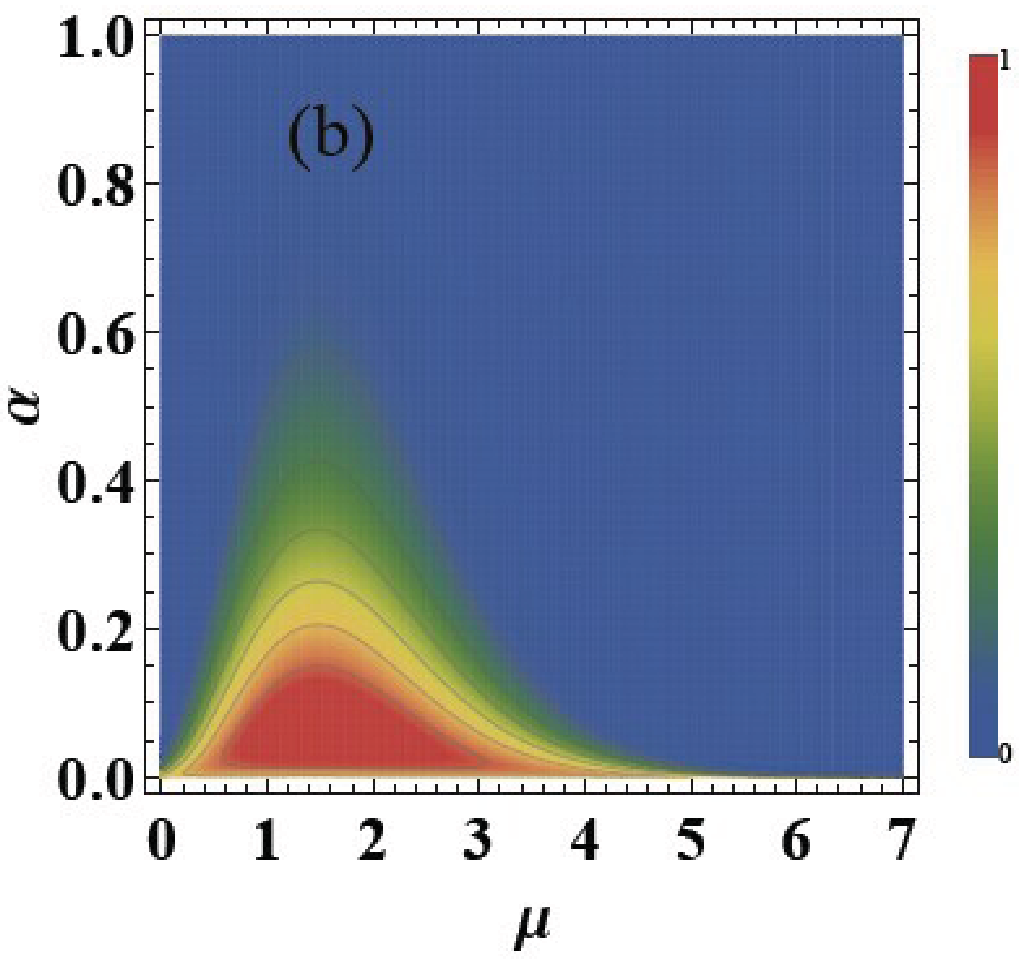}
\includegraphics*[bb=0 0 176 176,width=4cm, clip]{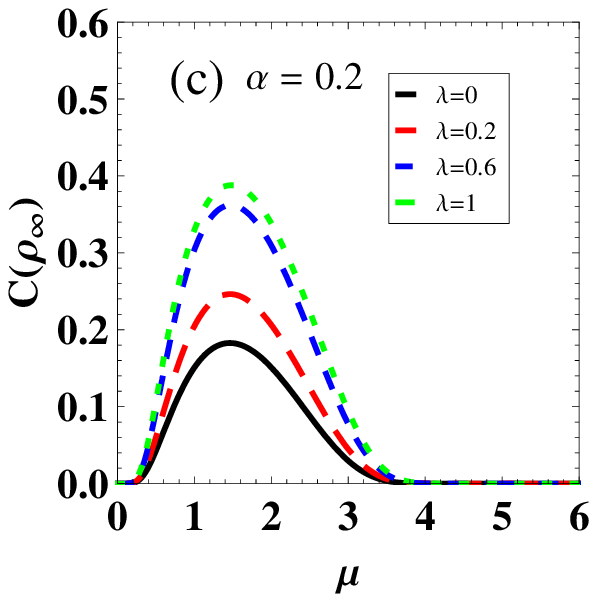}
\includegraphics*[bb=0 0 178 178,width=4cm, clip]{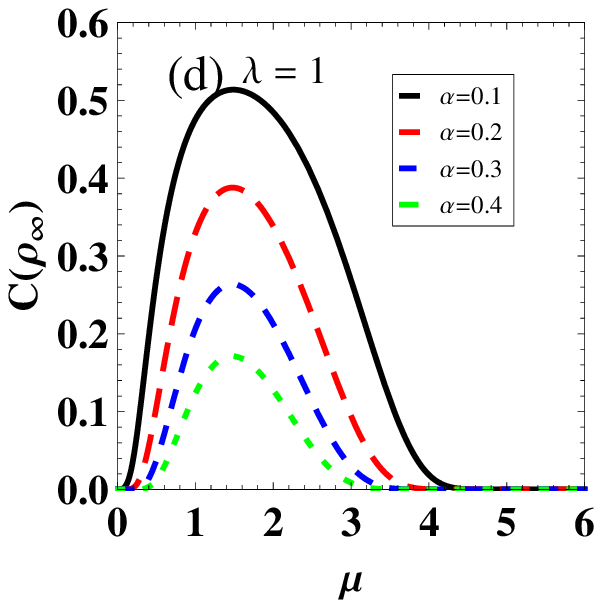}
\caption{(Color online) The stationary coherence of the qubit
quantified by the relative entropy of coherence $C(\rho_{\infty})$
as a function of the bath parameters $\alpha$ and $\mu$. (a) for the
uncorrelated initial qubit-bath state ($\lambda=0$); (b) for the
correlated initial qubit-bath state ($\lambda=1$); (c) $\alpha=0.2$;
(d) $\lambda=1$. Parameters are chosen as, $\upsilon=1.5$, and
$\omega_{c}=1$.}
\end{figure}

We shall examine the decoherence process where the initially
correlated qubit-bath state is in the form of Eq. (\ref{2}), with
$c_{e}=c_{g}=1/\sqrt{2}$. Then the initial coherence of the qubit
can be evaluated
$C(\rho^{t=0}_{\lambda})=\frac{1}{2}(1-\Upsilon_{\lambda}(0))\log_{2}[1-\Upsilon_{\lambda}(0)]+\frac{1}{2}(1+\Upsilon_{\lambda}(0))\log_{2}[1+\Upsilon_{\lambda}(0)]$,
with
$\Upsilon_{\lambda}(0)=C^{-1}_{\lambda}(1-\lambda+\lambda\exp[-\frac{1}{2}\Gamma[\upsilon]\omega^{\upsilon}_{c}])$.
At time $t=0$, in the case $\lambda=0$ the dephasing rate
$\Upsilon_{0}(0)=1$, while for the correlated initial state we can
obtain $0<\Upsilon_{\lambda}(0)<1$. So it means
$C(\rho^{\lambda\neq0}_{t=0})<C(\rho^{\lambda=0}_{t=0})$, that is to
say the initial correlation of the qubit-bath system can lead to
lower initial coherence of the qubit.

On the other hand, to clear the effect of the qubit-bath initial
correlation explicitly, we also perform the calculation for the
stationary value of coherence trapping in the long time limit. In
Fig. $1$, we show the stationary coherence $C(\rho_{\infty})$
between the initially uncorrelated $\lambda=0$ and correlated
$\lambda=1$ states as a function of the bath parameters $\alpha$ and
$\mu$. By comparing Figs. $1(a)$ and $1(b)$, it is clear that the
presence of the qubit-bath correlation in the initial state enlarges
the region for occurrence of coherence trapping. Moreover, by giving
the other parameters, Fig. $1(c)$ clearly shows that the larger
correlation parameter $\lambda$ leads to a more efficient coherence
trapping as the stationary coherence is higher than that obtained
from the initially uncorrelated qubit-bath state. Although the lower
initial coherence of the qubit can be induced by the correlation
parameter $\lambda$, the coherences of the bath subsystem and the
qubit-bath composite system would appear in the initial qubit-bath
state correspondingly. That is the main physical reason of the more
efficient coherence trapping of the qubit induced by the correlated
initial qubit-bath state. Additionally, from Fig. $1$ we also can
easily find that, the stronger coupling $\alpha$ of the qubit to
bath diminishes the stationary coherence in the long time limit. And
there exists an optimal value of the Ohmicity parameter
$\mu\doteq1.46$ of the bath maximizing the stationary coherence,
which is independent of the coupling constant $\alpha$ and the
correlation parameter $\lambda$, as shown in Figs. $1(c)$ and
$1(d)$.

Next, by choosing the optimal value $\mu=1.46$ of the super-Ohmic
bath, the influence of the parameters characterizing the initially
correlated state on coherence trapping is depicted in Fig. $2(a)$.
Two regions, the enhancing of coherence tapping (ECT) (i.e.
 $C(\rho^{\lambda\neq0}_{\infty})>C(\rho^{\lambda=0}_{\infty})=0.1827$ ) and the no-enhancing of
coherence trapping (No-ECT) (i.e.
$C(\rho^{\lambda\neq0}_{\infty}){\leq}C(\rho^{\lambda=0}_{\infty})$),
are acquired in the corresponding parameter planes. The dashed-white
line $C(\rho_{\infty})=0.1827$ is the dividing line between these
two regions. That is to say, not all but specific initial states
$|\xi_{\lambda}\rangle$ can lead to the enhancing coherence
trapping. The range of $\upsilon$ to gain the enhancing of coherence
trapping, would reduce as the correlation parameter $\lambda$
increasing, as shown in Fig. $2(b)$. So we conclude that, in order
to achieve the most efficient coherence in the long time limit, both
the optimal Ohmicity parameter $\mu$ and the optimal state
$|\xi_{\lambda}\rangle$ must be satisfied.

\begin{figure}[tbh]
\includegraphics*[bb=0 0 249 227,width=4.4cm, clip]{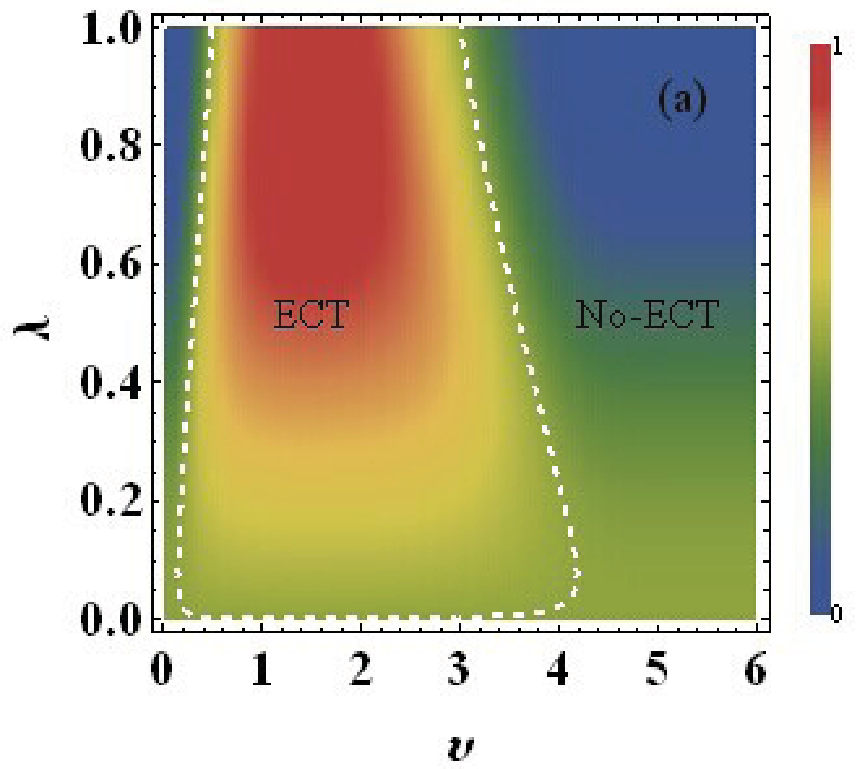}
\includegraphics*[bb=0 0 160 160,width=4.1cm, clip]{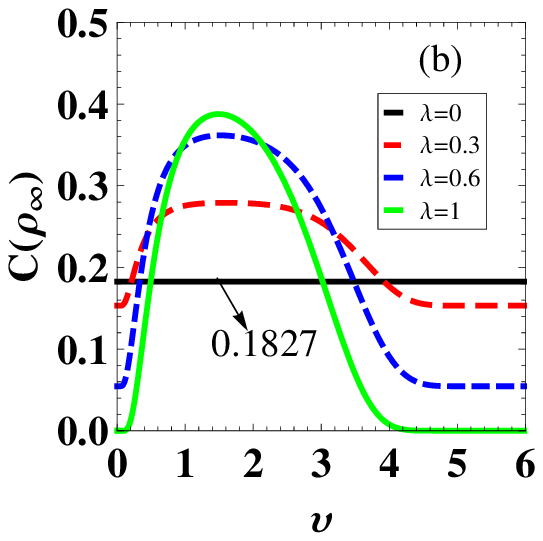}
\caption{(Color online) The stationary coherence of the qubit
quantified by the relative entropy of coherence $C(\rho_{\infty})$
as a function of the parameters for the initial qubit-bath state
$\lambda$ and $\upsilon$. The dashed-white line in $(a)$ means
$C(\rho_{\infty})=0.1827$, which is the dividing line between two
regions. Parameters are chosen as, $\alpha=0.2$, $\mu=1.46$, and
$\omega_{c}=1$. }
\end{figure}

{\it{Quantum evolution speed.}}
 Since the qubit would occur coherence trapping after a
finite time $t_{c}$ in the super-Ohmic bath, one may naturally
concern the evolution speed between the initial state
$\rho_{\lambda}(0)$ and the stationary coherence state
$\rho_{\lambda}(t_{c})$. The quantum speed of evolution from
$\rho_{\lambda}(0)$ to its target state $\rho_{\lambda}(t_{c})$ can
be characterized by QSL time \cite{33,34}. The definition of QSL
time between an arbitrary initially mixed state $\rho_{0}$ and its
target state $\rho_{\tau}$, governed by the master equation
$\dot{\rho}_{t}=L_{t}\rho_{t}$, with $L_{t}$ the positive generator
of the dynamical semigroup, is as follows \cite{34} $
\tau_{QSL}=\max\{\frac{1}{\overline{\sum^{n}_{i=1}\sigma_{i}\varrho_{i}}},\frac{1}{\overline{\sqrt{\sum^{n}_{i=1}\sigma^{2}_{i}}}}\}B(\rho_{0},\rho_{\tau}),
$ with $\overline{X}=\tau^{-1}\int^{\tau}_{0}Xdt$,
$B(\rho_{0},\rho_{\tau})=|tr(\rho_{0}\rho_{\tau})-tr(\rho^{2}_{0})|$
denotes a metric on the space of the initial state $\rho_{0}$ and
the target state $\rho_{\tau}$ via the so-called relative purity,
and $\sigma_{i}$ are the singular values of $\dot{\rho}_{t}$ and
$\varrho_{i}$ those of the initial mixed state $\rho_{0}$.  The
above expression of $\tau_{QSL}$ can effectually define the minimal
evolution time for arbitrary initial states, and also be used to
assess quantum evolution speed of open quantum system.
\begin{figure}[tbh]
\includegraphics*[bb=0 0 188 188,width=4cm, clip]{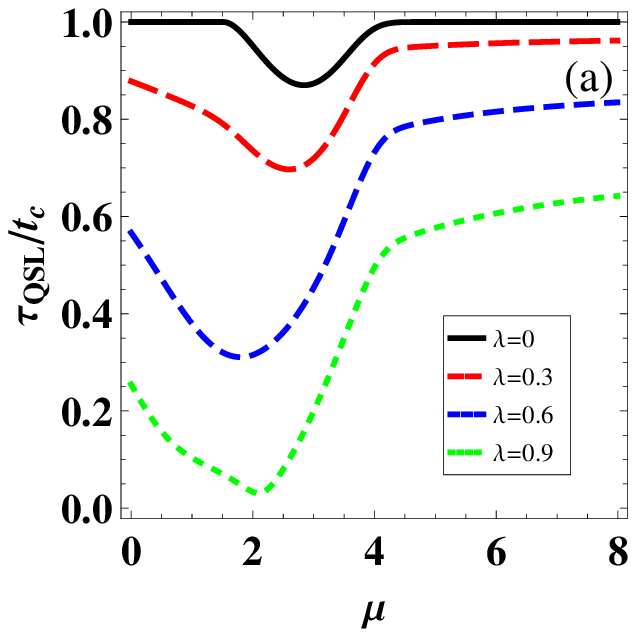}
\includegraphics*[bb=0 0 191 191,width=4cm, clip]{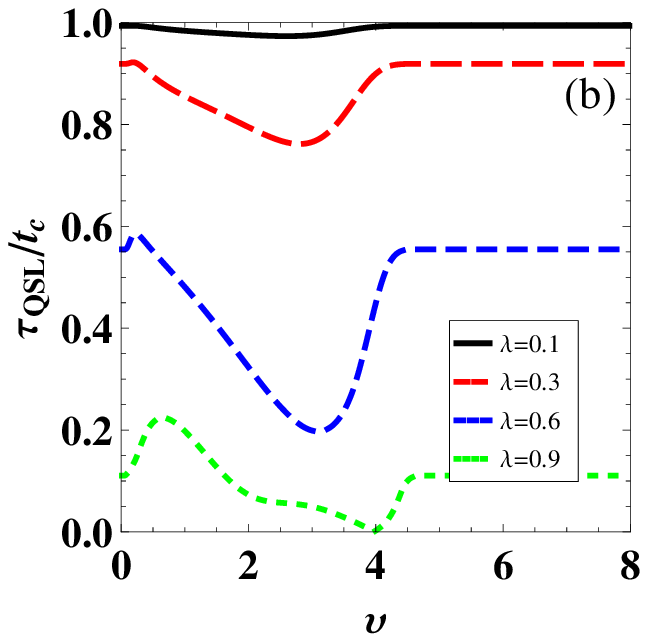}
\includegraphics*[bb=0 0 326 326,width=4.5cm, clip]{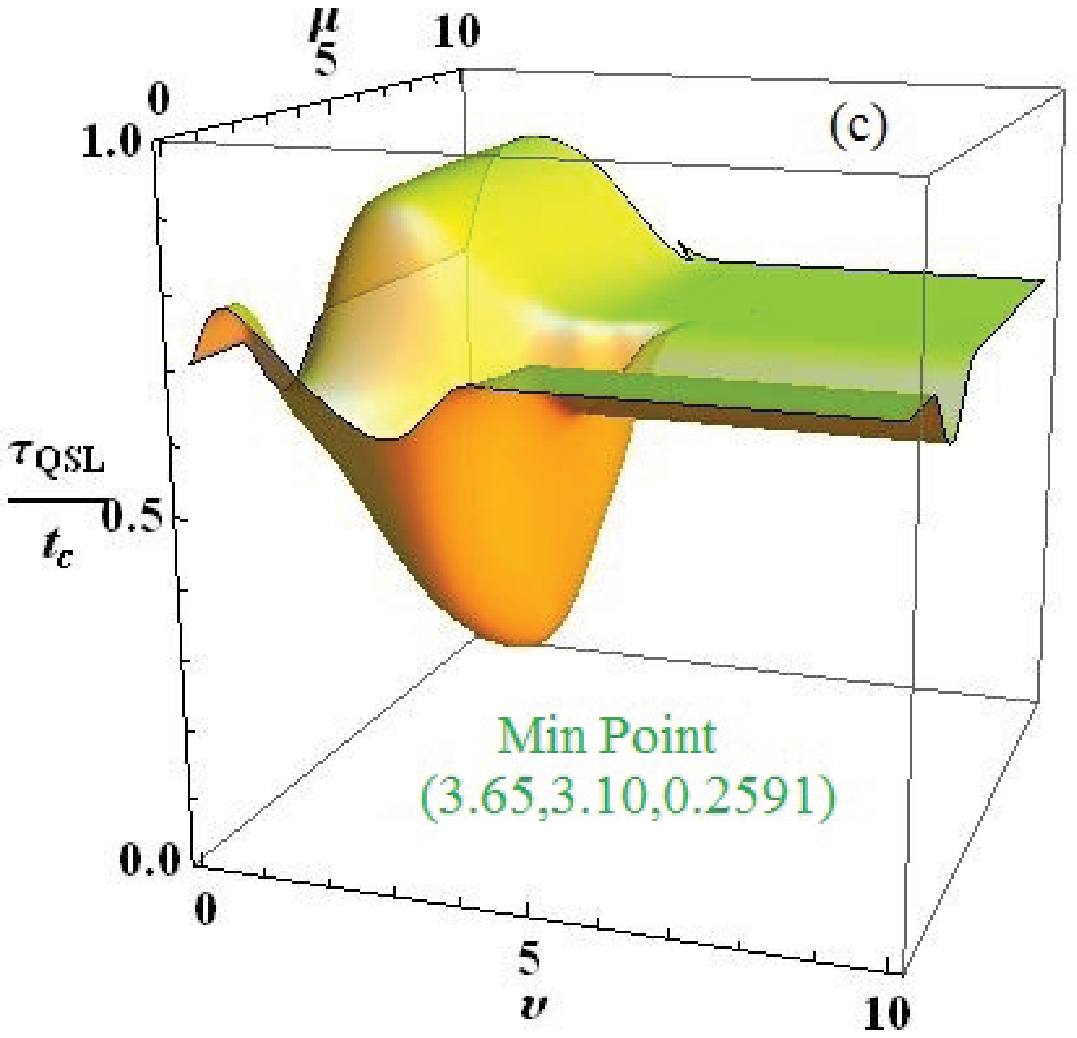}
\caption{(Color online) The QSL time for the evolution from the
initial state $\rho_{\lambda}(0)$ to the stationary coherence state
$\rho_{\lambda}(t_{c})$, quantified by $\tau_{QSL}/t_{c}$ as a
function of the parameters $\mu$ and $\upsilon$, with
$\omega_{c}=1$. Parameters are chosen as, (a) $\alpha=0.2$,
$\upsilon=2$; (b) $\alpha=0.2$, $\mu=1.46$; (c) $\alpha=0.2$,
$\lambda=0.5$. }
\end{figure}

Here, we also consider the weights $c_{e}=c_{g}=1/\sqrt{2}$ in the
initially correlated qubit-bath state in Eq. (\ref{2}). Then the QSL
time for the qubit initial state $\rho_{\lambda}(0)$ to the
stationary coherence state $\rho_{\lambda}(t_{c})$,  can be
calculated
$\tau_{QSL}/t_{c}=|\Upsilon_{\lambda}(0)[\Upsilon_{\lambda}(t_{c})-\Upsilon_{\lambda}(0)]|/\int^{t_{c}}_{0}|\dot{\Upsilon}_{\lambda}(t)|dt.$
In Figs. $3(a)$ and $3(b)$, we demonstrate how the QSL time for
evolution from $\rho_{\lambda}(0)$ to $\rho_{\lambda}(t_{c})$ can
depend on the parameters $\mu$ and $\upsilon$, with different
selected correlation parameter $\lambda$. Firstly, it is clear that
the initial qubit-bath correlation can reduce the QSL time as the
value of $\lambda$ increasing. That is to say, the evolution from
the initial coherence state to the stationary coherence state, can
be speeded up by the initial correlation in the qubit-bath state.
And then, another remarkable feature can be acquired: There exist
the optimal Ohmicity parameter $\mu$ or the parameter $\upsilon$ of
$|\xi_{f}\rangle$, which can induce the minimum value of QSL time.
And the optimal parameters $\mu$ or $\upsilon$ are dependent of the
correlation parameter $\lambda$. In Fig. $3(a)$, when $\upsilon=2$,
the optimal Ohmicity parameter $\mu\doteq2.84,2.60,1.80,2.09$ for
$\lambda=0,0.3,0.6,0.9$, respectively. By choosing $\mu=1.46$, as
shown in Fig. $3(b)$, the optimal parameter for the initial bath
state $|\xi_{f}\rangle$ can be obtain
$\upsilon\doteq2.61,2.80,3.09,3.97$ for $\lambda=0.1,0.3,0.6,0.9$,
respectively.

Furthermore, since both the Ohmicity parameter $\mu$ and the
parameter $\upsilon$ of $|\xi_{f}\rangle$ can bring about the
minimum QSL time, in the following we would seek the optimal
condition $(\upsilon,\mu)$ on the maximal evolution speed of the
qubit. Fig. $3(c)$ shows QSL time for $\rho_{\lambda}(0)$ to
$\rho_{\lambda}(t_{c})$ as a function of $\mu$ and $\upsilon$. By a
given correlation parameter $\lambda=0.5$, we observe that, the
minimum QSL time can only appear in the region $(\upsilon<5,\mu<4)$.
And the optimal values $(\upsilon=3.65,\mu=3.10)$ which lead to the
minimum QSL time $\tau^{min}_{QSL}/t_{c}=0.2591$, can be found by
accurate numerical calculation. This can be understand that, in
order to speed up the evolution speed of the qubit, the Ohmicity
parameter $\mu$ and the parameter $\upsilon$ of $|\xi_{f}\rangle$
should be optimized. Combined with the above section about coherence
trapping, the aim to make the qubit trap in a higher stationary
coherence state with the maximal evolution speed, can be attained by
choosing the optimal parameters of the initial qubit-bath state
($\lambda,\upsilon$) and the bath spectral density function ($\mu$).

{\it{Conclusion.}} In summary, we studied intriguing features of
coherence trapping of a qubit with a zero-temperature structured
bath by considering the initial qubit-bath correlation. The initial
qubit-bath correlation not only leads to a more efficient coherence
trapping, but also speeds up the evolution for the occurrence of
coherence trapping. Moreover, both the maximum stationary coherence
in the long time limit and the minimum QSL time from the initial
state to the stationary coherence state, can be acquired by
optimizing the parameters of the initially correlated qubit-bath
state and the bath spectral density. This physical mechanism leading
quickly to a higher stationary coherence would play an important
role for implementing quantum simulators \cite{43} and quantum
information processors \cite{44}. It is worth pointing out that the
non-Markovian effects may not monotonically cause the acceleration
of the system evolution in the super-Ohmic bath, as shown in Fig.
$3(a)$. This is clearly different from the main result in the damped
Jaynes-Cummings model \cite{33}, which shows that the evolution
speed can be monotonically increased by non-Markovian effects. So
the specific interplay between the evolution speed of the system and
the bath non-Markovian effects should be studied under different
circumstances. Experimentally, the coherence trapping can be
demonstrated by qubit-bath systems like optics \cite{20}, trapped
ions \cite{trapped-ions} and superconducting qubit
\cite{44,supercond1,supercond2}.

{\it{Acknowledgements.}}
 This work was supported by the National Natural Science
Foundation of China under grant Nos. 11304179, 11247240, 11175248,
61178012, 11204156, the Specialized Research Fund for the Doctoral
Program of Higher Education under grant Nos. 20133705110001,
20123705120002, the Provincial Natural Science Foundation of
Shandong under grant Nos. ZR2012FQ024, ZR2014AP009, and XDB grants
of Chinese Academy of Sciences.


\begin{thebibliography}{99}
\bibitem{1}L. Mandel, and E. Wolf, \textit{Optical Coherence and Quantum Optics} (Cambridge University
Press, Cambridge, 2008).
\bibitem{2}L. M. K. Vandershypen, and I. L. Chuang, Rev. Mod. Phys. \textbf{76}, 1037 (2004).
\bibitem{3}H. P. Breuer, and F. Petruccione, \textit{The Theory of Open Quantum Systems} (Oxford University
Press, Oxford, 2007).
\bibitem{4}H. Lee, Y. C. Cheng, and G. R. Fleming, Science \textbf{316}, 1462 (2007).
\bibitem{5}L. S. Cederbaum, E. Gindensperger, and I. Burghardt, Phys. Rev. Lett. \textbf{94}, 113003 (2005).
\bibitem{6}P. Rebentrost, and A. Aspuru-Guzik, J. Chem. Phys. \textbf{134}, 101103 (2011).
\bibitem{7}J. G. Tony, T. J. G. Apollaro, C. Di Franco, F. Plastina, and M. Paternostro, Phys. Rev. A \textbf{83},
032103 (2011).
\bibitem{8}S. F. Huelga, A. Rivas, and M. B. Plenio, Phys. Rev. Lett. \textbf{108}, 160402 (2012).
\bibitem{9}A. W. Chin, J. Prior, R. Rosenbach, F. Caycedo-Soler, S. F. Huelga, and M. B. Plenio, Nat. Phys. \textbf{9}, 113 (2013).
\bibitem{10}C. Addis, G. Brebner, P. Haikka, and S. Maniscalco, Phys. Rev. A \textbf{89}, 024101 (2014).
\bibitem{11}C. F. Li, J. S. Tang, Y. L. Li and G. C. Guo, Phys. Rev. A \textbf{83}, 064102 (2011).
\bibitem{12}Y. J. Zhang, X. B. Zou, Y. J. Xia and G. C. Guo, Phys. Rev. A \textbf{82}, 022108 (2010).
\bibitem{13}W. M. Zhang, P. Y. Lo, H. N. Xiong, M. W. Y. Yu, and F. Nori, Phys. Rev. Lett. \textbf{109}, 170402 (2012).
\bibitem{14}A. G. Dijkstra, and Y. Tanimura, Phys. Rev. Lett. \textbf{104}, 250401 (2010).
\bibitem{15}A. Shabani, and D. A. Lidar, Phys. Rev. Lett. \textbf{102}, 100402 (2009).
\bibitem{015}A. Z. Chaudhry, and J. B. Gong, Phys. Rev. A \textbf{88}, 052107 (2013).
\bibitem{16}A. Smirne, D. Brivio, S. Cialdi, B. Vacchini, and M. G. A. Paris, Phys. Rev. A \textbf{84}, 032112 (2011).
\bibitem{17}J. Dajka, and J. {\L}uczka, Phys. Rev. A \textbf{82}, 012341 (2010).
\bibitem{18}H. P. Breuer, E. M. Laine, and J. Piilo, Phys. Rev. Lett. \textbf{103}, 210401 (2009).
\bibitem{19}A. Rivas, S. F. Huelga, and M. B. Plenio, Phys. Rev. Lett. \textbf{105}, 050403 (2010).
\bibitem{20}B. H. Liu, L. Li, Y. F. Huang, C. F. Li, G. C. Guo, E. M. Laine, H. P. Breuer, and J. Piilo, Nat. Phys. \textbf{7}, 931 (2011).
\bibitem{21}E. M. Laine, J. Piilo, and H. P. Breuer,  Europhys. Lett. \textbf{92}, 60010 (2010).
\bibitem{021}M. Nielsen, and I. Chuang, \textit{Quantum Computation and Quantum Communication} (Cambridge University
Press, Cambridge, 2000).
\bibitem{0021}A. D. Cimmarusti, C. A. Schroeder, B. D. Patterson, L. A. Orozco, P. Barberis-Blostein, and H. J. Carmichael, New J. Phys. \textbf{15}, 013017 (2013).
\bibitem{022}G. C. Hegerfeldt, Phys. Rev. Lett. \textbf{111}, 260501 (2013).
\bibitem{0022}G. C. Hegerfeldt, Phys. Rev. A \textbf{90}, 032110 (2014).
\bibitem{22}L. Mandelstam, and I. Tamm, J. Phys. (USSR) \textbf{9}, 249-254 (1945).
\bibitem{23}J. Anandan, and Y. Aharonov, Phys. Rev. Lett. \textbf{65}, 1697-1700 (1990).
\bibitem{24}L. B. Levitin, and T. Toffoli, Phys. Rev. Lett. \textbf{103}, 160502 (2009).
\bibitem{25} V. Giovannetti, S. Lloyd, and L. Maccone, Phys. Rev. A \textbf{67},
052109 (2003).
\bibitem{26}P. J. Jones, and P. Kok, Phys. Rev. A \textbf{82}, 022107 (2010).
\bibitem{27}M. Zwierz, Phys. Rev. A \textbf{86}, 016101 (2012).
\bibitem{28}S. Deffner, and E. Lutz, J. Phys. A: Math. Theor. \textbf{46} 335302(2013).
\bibitem{29}P. Pfeifer, Phys. Rev. Lett. \textbf{70}, 3365 (1993).
\bibitem{30} P. Pfeifer, and J. Fr$\ddot{o}$hlich, Rev. Mod. Phys. \textbf{67}, 759(1995).
\bibitem{31}M. M. Taddei, B. M. Escher, L. Davidovich, and R. L. de Matos Filho, Phys. Rev. Lett. \textbf{110}, 050402 (2013).
\bibitem{32}A. del Campo, I. L. Egusquiza, M. B. Plenio, and S. F. Huelga, Phys. Rev. Lett. \textbf{110}, 050403 (2013).
\bibitem{33}S. Deffner, and E. Lutz, Phys. Rev. Lett. \textbf{111}, 010402 (2013).
\bibitem{34}Y. J. Zhang, W. Han, Y. J. Xia, J. P. Cao, and H. Fan, Sci. Rep. \textbf{4}, 4890 (2014).
\bibitem{35}Z. Y. Xu, S. Luo, W. L. Yang, C. Liu, and S. Q. Zhu, Phys. Rev. A \textbf{89}, 012307 (2014).
\bibitem{38}J. Dajka, and J. {\L}uczka, Phys. Rev. A \textbf{77}, 062303 (2008).
\bibitem{39}J. Dajka, M. Mierzejewski, and J. {\L}uczka, Phys. Rev. A \textbf{79}, 012104 (2009).
\bibitem{40}T. Baumgratz, M. Cramer, and M. B. Plenio, Phys. Rev. Lett. \textbf{113}, 140401 (2014).
\bibitem{41}J. {\AA}berg, Phys. Rev. Lett. \textbf{113}, 150402 (2014).
\bibitem{42}D. Girolami, Phys. Rev. Lett. \textbf{113}, 170401 (2014).
\bibitem{43}J. I. Cirac, and P. Zoller, Nat. Phys. \textbf{8}, 264 (2012).
\bibitem{44}I. M. Georgescu, S. Ashhab, and F. Nori, Rev. Mod. Phys. \textbf{86}, 153 (2014).

\bibitem{trapped-ions}J. T. Barreiro, M. Muller, P. Schindler, D. Nigg, T. Monz, M. Chwalla,
M. Hennrich, C. F. Roos, P. Zoller, and R. Blatt, Nature {\bf 470}, 486 (2011).

\bibitem{supercond1}P. Forn-Diaz, J. Lisenfeld, D. Marcos, J. J. Garcia-Ripoll, E. Solano,
C. Harmans, and J. E. Mooij, Phys. Rev. Lett. {\bf 105}, 237001 (2010).

\bibitem{supercond2}Q. T. Xie, S. Cui, J. P. Cao, A. Luigi, and H. Fan,
 Phys. Rev. X {\bf 4}, 021046 (2014).

\end{thebibliography}
\end{document}